%% file: panel.tex
\documentclass[preprint]{imsart}

\RequirePackage[OT1]{fontenc}
\usepackage{amsthm,amsmath,amsthm}
\usepackage[authoryear]{natbib}

\usepackage{graphicx,amsfonts,subcaption,bbm}
\usepackage{rotating,tikz,mathtools,url,lscape,longtable,verbatim}

\RequirePackage[colorlinks,citecolor=blue,urlcolor=blue]{hyperref}

\newcommand\independent{\protect\mathpalette{\protect\independenT}{\perp}}
\def\independenT#1#2{\mathrel{\rlap{$#1#2$}\mkern2mu{#1#2}}}

\theoremstyle{plain}

\startlocaldefs
\numberwithin{equation}{section}
\theoremstyle{plain}

\endlocaldefs

\begin{document}

\begin{frontmatter}

\title{Assessing the causal effect of binary interventions from observational panel data with few treated units\\ \normalsize{\today}}
\runtitle{Causal inference with observational panel data}


\author{\fnms{Pantelis} \snm{Samartsidis}\corref{}\ead[label=e1]{pantelis.samartsidis@mrc-bsu.cam.ac.uk}},
\author{\fnms{Shaun R.} \snm{Seaman}\ead[label=e2]{shaun.seaman@mrc-bsu.cam.ac.uk}},
\author{\fnms{Anne M.} \snm{Presanis}\ead[label=e5]{anne.presanis@mrc-bsu.cam.ac.uk}},
\author{\fnms{Matthew} \snm{Hickman}\ead[label=e3]{matthew.hickman@bristol.ac.uk}}
\and
\author{\fnms{Daniela} \snm{De Angelis}\ead[label=e4]{daniela.deangelis@mrc-bsu.cam.ac.uk}}

\address{MRC Biostatistics Unit, University of Cambridge\\ University Forvie Site, Robinson Way, Cambridge CB2 0SR, UK\\ \printead{e1,e2}\\ \printead{e5,e4}}
\address{Population Health Sciences, Bristol Medical School, University of Bristol,\\ Bristol BS8 2PS, UK\\ \printead{e3}}

\affiliation{MRC Biostatistics Unit, University of Cambridge}
\runauthor{P. Samartsidis et al}

\begin{abstract}
Researchers are often challenged with assessing the impact of an intervention on an outcome of interest in situations where the intervention is non-randomised, the intervention is only applied to one or few units, the intervention is binary, and outcome measurements are available at multiple time points. In this paper, we review existing methods for causal inference in these situations. We detail the assumptions underlying each method, emphasize connections between the different approaches and provide guidelines regarding their practical implementation. Several open problems are identified thus highlighting the need for future research.  
\end{abstract}



\end{frontmatter}

\input{introduction.tex}

\input{preliminaries.tex}

\input{meth.tex}

\input{comparison.tex}

\input{real.tex}

\input{discussion.tex}

\section*{Acknowledgements}
We acknowledge funding and support from NIHR Health Protection Unit on Evaluation of Interventions ((PS, MH, DDA), Medical Research Council grants MC\_UU\_00002/10 (SRS) and MC\_UU\_00002/11 (DDA, AMP), Public Health England (DDA), and NIHR PGfAR RP-PG-0616-20008 (EPIToPe, PS, MH, DDA). The views expressed are those of the authors and not necessarily those of the NHS, the NIHR or the Department of Health.
 
\bibliographystyle{imsart-nameyear}
\bibliography{causal,did,sc,factors}

\newpage
\input{appendix.tex}

\end{document}

%% file: introduction.tex
\section{Introduction}\label{sec:intro}

Evaluation of the causal effect of an intervention (\textit{e.g}.\ a newly introduced policy, a novel experimental practice or an unexpected event) on an outcome of interest is a problem frequently encountered in several fields of scientific research. 
These include economics \citep{Angrist2009,Imbens2009}, epidemiology and public health  \citep{Rothman2005,Glass2013}, management \citep{Antonakis2010}, marketing \citep{Rubin2006,Varian2016}, and political sciences \citep{Keele2015}. 

Researchers are often interested in assessing the impact of an intervention (occasionally referred to as treatment henceforth) in situations where: i) the data are observational, \textit{i.e.}\ the allocation of the sample units to the intervention and control groups is not randomised, but instead determined by factors that confound the association between the indicator of intervention and the outcome of interest; ii) the intervention is binary, \textit{i.e.}\ sample units cannot receive the interventions at varying intensities; iii) only one or a small number of units are treated; and iv) \textcolor{black}{at each of a set of time points, before and after the time at which the intervention is introduced, the outcome is measured on every sampled unit, thus giving rise to panel data.}


Several statistical methods for causal inference in this setting have been developed to account for the special characteristics of the data: the presence of (likely unobserved) confounders, the existence of temporal trends in the outcome and the limited number of sample units to which the intervention is given. 
\textcolor{black}{
In this paper we review the existing literature, motivated by recent methodological developments and the increasing application of these methods to real-life problems. 
Since the existing literature comes from a wide range of research disciplines, our focus is on unifying the various methods under a common terminology and notation, appropriate to a statistical audience. 
Further, we draw connections between various methods and point out issues related to their practical implementation. 
Finally, we suggest some possible directions for future research.  
}

We focus on four classes of methods: \textit{difference-in-differences}, \textit{latent factor models}, \textit{\textcolor{black}{synthetic control-type methods}} and the \textit{causal impact} method. 
Excluded from our review are propensity score methods \citep{Rosenbaum1983} (see \citet{Austin2011} for a recent review), because the \textcolor{black}{small number of treated units} does not allow accurate estimation of the parameters of a propensity score model, and the interrupted time series method \citep{Bernal2016}, because it does not use data on the units that do not receive the intervention.

This manuscript is structured as follows. 
In Section \ref{sec:background} we define notation, describe the causal framework underlying the methods and introduce the illustrative example. 
Section \ref{sec:met} presents the four \textcolor{black}{classes of methods}. 
\textcolor{black}{
Section \ref{sec:uncertainty} is about quantification of uncertainty and hypothesis testing. 
Section \ref{sec:practice} discusses issues related to practical implementation. 
Sections \ref{sec:real1} and \ref{sec:discussion} contain an applications to real data and a discussion, respectively. 
Finally, in Section \ref{sec:future} we highlight some remaining problems in the field.
}

%% file: preliminaries.tex
\section{Preliminaries}\label{sec:background}
\subsection{Notation} \label{sec:notation}
Let $i=1,\dots ,n$ index the entities (e.g.\ hospitals or general practices) for which the outcome of interest is observed: henceforth, we refer to these entities as units. 
For unit $i$ we have measurements $\boldsymbol{y}_{i\cdot}=(y_{i1},\dots,y_{iT})^\top$, where $t$ indexes time. 
We let $\boldsymbol{y}_{\cdot t}=(y_{1t},\dots,y_{nt})^\top$ denote the vector containing the set of $n$ observations at time $t$ and $\boldsymbol{y}_{i,t_1:t_2}=(y_{it_1},\dots,y_{it_2})^\top$ denote the measurements on unit $i$ from time $t_1$ to time $t_2$ ($t_1\leq t_2$). 
Throughout, we assume that $y_{it}$ is univariate.

Let $d_{it}=1$ if unit $i$ receives the intervention at or before time $t$, and $d_{it}=0$ otherwise. 
Let $\boldsymbol{d}_i=(d_{i1},\dots,d_{iT})^\top$.
Of the $n$ units, the first $n_1$ remain untreated for the entire study period. 
We call these the \textit{controls}. 
For the $n_2$ \textit{treated} units, there is a time $T_{1}$ ($1<T_1<T$) immediately after which the intervention is applied.  
We assume that all treated units receive the intervention at the same time. 
Hence $d_{it}=1$ if $i>n_1$ and $t>T_1$, and $d_{it}=0$ otherwise. 
We make this assumption to simplify the notation, but the methods we describe can be easily extended to allow for different treatment times. 
The number of post-intervention observation times is denoted by $T_2=T-T_1$. 
For each unit and time we may also observe a set of $K$ covariates $\boldsymbol{x}_{it}=(x_{it1},\dots x_{it K})^\top$.

\subsection{Potential outcomes}
We adopt the \textit{potential outcomes} framework \citep{Rubin1974,Rubin1990}, also known as the \textit{Rubin causal model} \citep[RCM]{Holland1986}. 
\textcolor{black}{
Under this model, for each treated unit ($i>n_1$) and post-intervention time ($t>T_1$) there are two potential outcomes, $y_{it}^{(0)}$ and $y_{it}^{(1)}$: $y_{it}^{(0)}$ represents the outcome that would be observed if intervention were not applied, and $y_{it}^{(1)}$ is the outcome that would be observed if the intervention were applied. 
We only observe $y_{it}^{(1)}$, \textit{i.e}., $y_{it}=y_{it}^{(1)}$. 
For the control units ($i\leq n_1$) at any time $t$ and the treated units ($i>n_1$) at a pre-intervention time ($t\leq T_1$), we only define $y_{it}^{(0)}$ and $y_{it}^{(0)}=y_{it}$ is observed.
}

The RCM allows the effect of intervention unit $i$ ($i>n_1$) at time $t$ ($t>T_1$) to be expressed as $\tau_{it} = y_{it}^{(1)} - y_{it}^{(0)}$. 
Estimation of $\tau_{it}$ is complicated by the fact that $y_{it}^{(0)}$ is not observed. 
In order to estimate $\tau_{it}$ from the observed data, it is necessary to make identifying assumptions \citep{Morgan2007,Keele2013}. 
For the methods considered in this paper, these assumptions allow the unobserved \textit{counterfactual} outcomes $y_{it}^{(0)}$ of treated units in the post-intervention period (\textit{i.e}.\ for $i>n_1$ and $t> T_1$) to be predicted using the observed outcomes on control and treated units. Denote these predictions as $\hat{y}_{it}^{(0)}$. 
The intervention effect $\tau_{it}$ can then be estimated as $\hat{\tau}_{it}=y_{it}-\hat{y}_{it}^{(0)}$. 

\subsection{An illustrative example}\label{sec:germany}
\begin{figure}[h]
        \centering
        \includegraphics[scale=0.45]{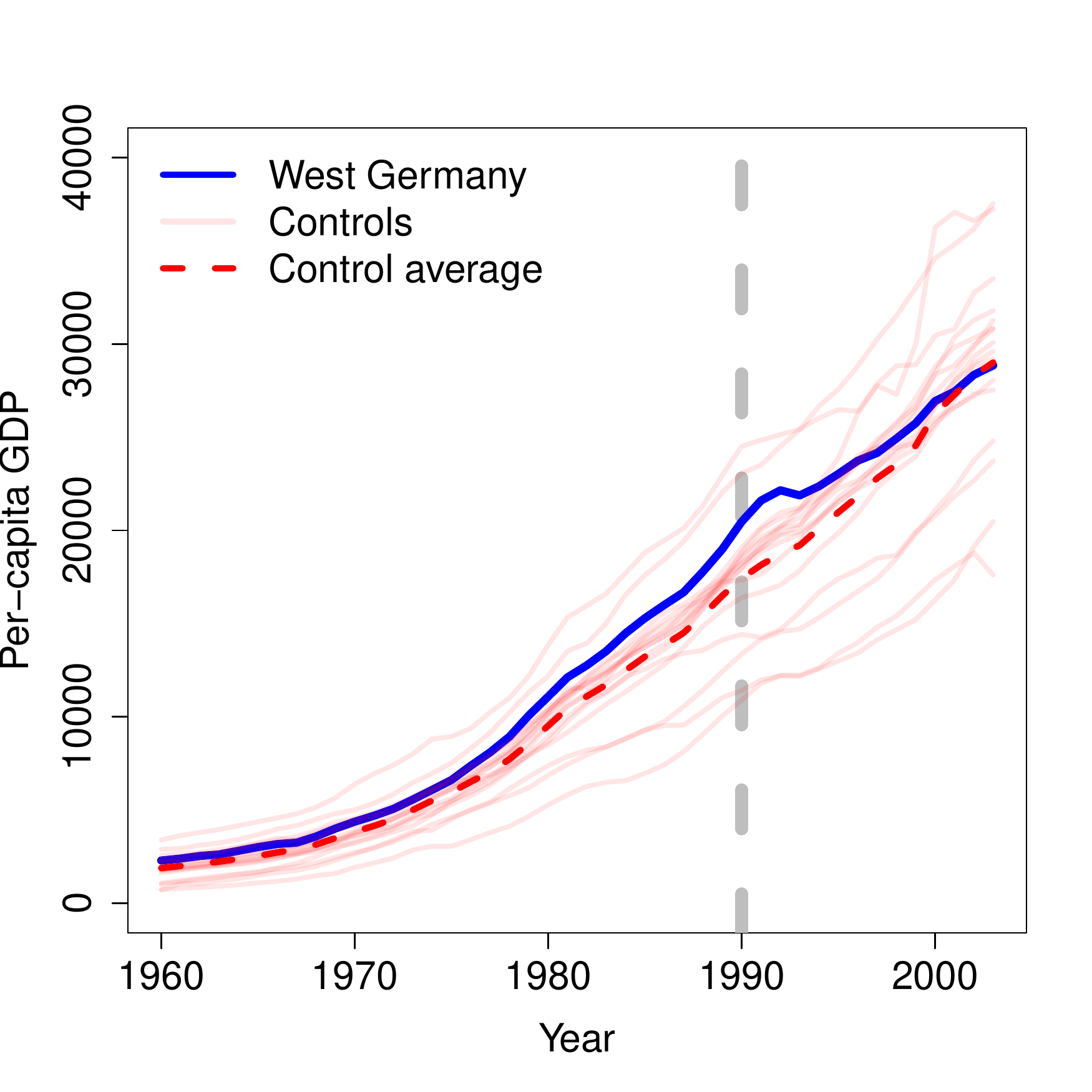}
        \vspace{-0.21in}
      \caption{Time series plot of the German reunification data. The values on the $y$-axis represent per-capita GDP measured in U.S. dollars. West Germany's per-capita GDP is shown in blue; the data on control units (16 other OECD countries) are shown in light red; the dashed red line represents the average GDP of the control units. The dashed gray line indicates 1990, the year of reunification.}      
      \label{fig:germany}
 \end{figure}

As our illustrative example we use data from \citet{Abadie2015} who investigated the effect that West Germany's reunification with East Germany in 1990 had on the economic growth of the former. 
To do so, they compared West Germany's annual per-capita GDP (the outcome variable) to its counterfactual GDP (i.e.\ its GDP had reunification not taken place), which they predicted based on annual per-capita GDP data from $n_1=16$ member countries of the Organisation for Economic Co-operation and Development (OECD) (none of which underwent reunification, and so are `control units'). 
The authors used data from 1960-2003 and hence there are $T_1=30$ pre-intervention and $T_2=13$ post-intervention time points. 
\textcolor{black}{
Figure \ref{fig:germany} shows the time-series of the outcome on all 17 units. 
In Section \ref{sec:real1}, we analyse this dataset using the methods reviewed in this article.    
}

%% file: meth.tex
\section{Estimation methods} \label{sec:met}

In this section, we \textcolor{black}{review four classes of methods} for predicting the counterfactual treatment-free outcomes $y_{it}^{(0)}$ of the treated units at post-intervention times, needed to calculate $\hat{\tau}_{it}$. 
\textcolor{black}{
Here we focus on the intuition and the assumptions underlying each method, and report results on theoretical properties of unbiasedness and consistency in Appendix \ref{sec:theory}. 
For full technical details of each approach, the reader is directed to the original publications.
}

\subsection{Difference-in-differences}\label{sec:did}
Early works \citep{Ashenfelter1978,Ashenfelter1985,Card1994} used so-called difference-in-differences (DID) models to compare two time periods (pre versus post-intervention). 
The identifying assumption in DID models is that the average outcomes of control and treated units in the absence of an intervention would follow \textit{parallel trends} over time \citep{Abadie2005}. 

\begin{figure}[h]
        \centering
        \includegraphics[scale=0.40]{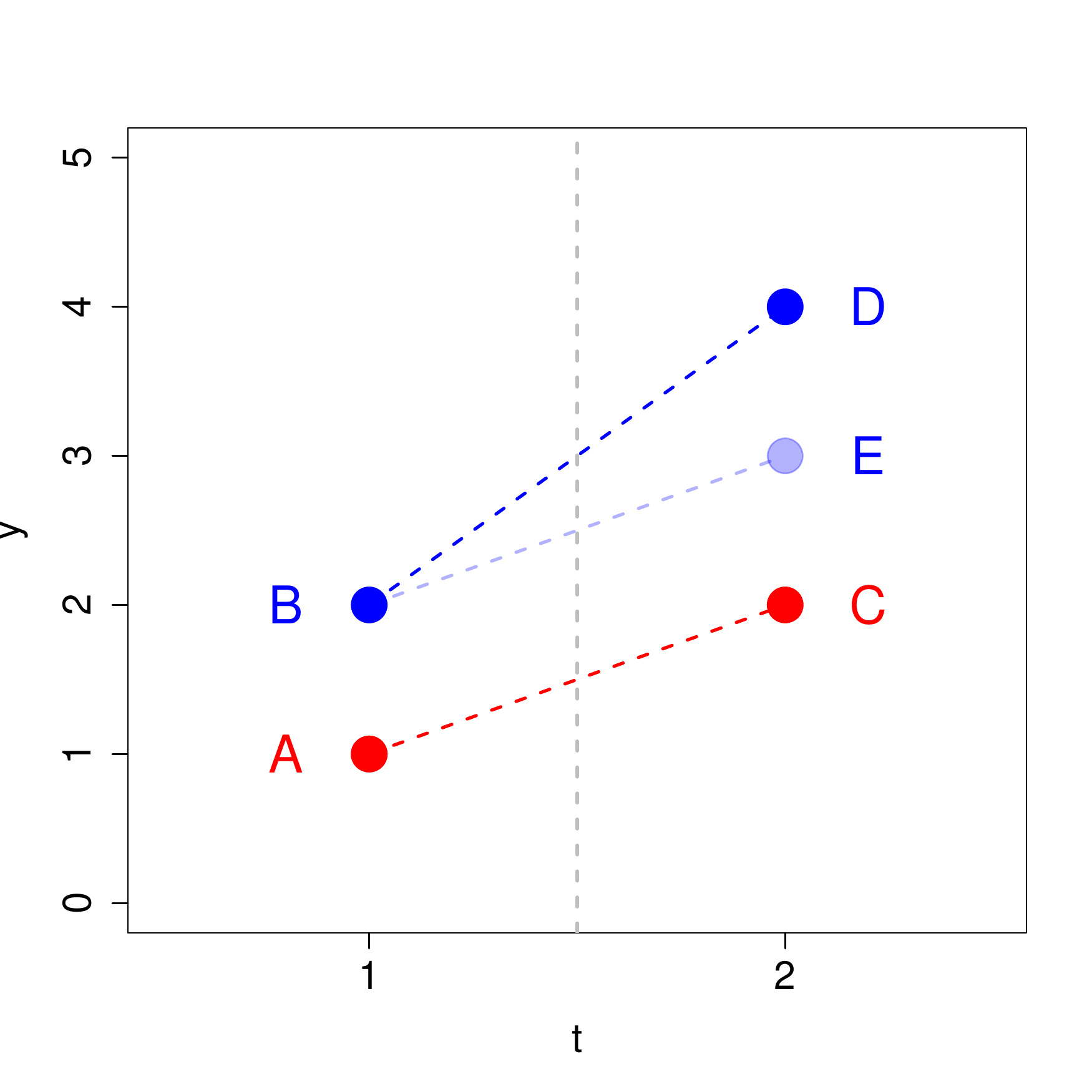}
        \vspace{-0.21in}
      \caption{Graphical illustration of the difference-in-differences method.} 
      \label{fig:did}
 \end{figure}

Figure \ref{fig:did} is a graphical representation of the basic DID method \textcolor{black}{for a single control and single treated unit. 
The four points A-D on the graph represent the control (A) and treated (B) units at $t=1$, and the control (C) and treated (D) units at $t=2$. 
Under the parallel trends assumption, the difference between the outcome of the treated unit and of the control unit would be constant over time in the absence of intervention. 
The counterfactual outcome for the treated unit at the post-intervention time can then be predicted as point E in Figure \ref{fig:did}. 
Letting $y_A,y_B,y_C,y_D$ and $y_E$ denote the $y$-values corresponding to the points A, B, C, D and E in Figure \ref{fig:did}, respectively, the estimated effect of the intervention is
\begin{eqnarray}\label{eq:dida}\nonumber
 \hat\tau_{22}& = & y_D - y_E\\ \nonumber
 & = & y_D - \left\{y_C +\left\{y_B-y_A \right\} \right\} \\ \nonumber
 & = & \left\{y_D - y_C\right\} - \left\{y_B-y_A\right\},
\end{eqnarray}
\textit{i.e.} the difference (after versus before) of the differences between the two units. 
The same method can be used when multiple time points \textcolor{black}{and multiple control units} are available.}

A commonly used solution to adjust for the effect of covariates $\boldsymbol{x}_{it}$\footnote{\textcolor{black}{\citet{Blundell2004} and \citet{Abadie2005} propose alternative DID estimators that can account for the effect of covariates. However, these methods are not suitable when only a small number of units are treated and hence are not reviewed in this article.}} is to specify a parametric linear DID model for the observed outcome $y_{it}$ \citep{Angrist2009,Jones2011}
\begin{eqnarray}\label{eq:didb}\nonumber
 y_{it}&=&y_{it}^{(0)}+\tau_{it} d_{it} \\
 y_{it}^{(0)}&=&\boldsymbol{x}_{it}^\top\boldsymbol\theta+\kappa_i+\mu_t+\varepsilon_{it},
\end{eqnarray}
where $\boldsymbol{\theta}$ is a vector of regression coefficients, $\kappa_i$ is an (unknown) fixed effect of unit $i$, $\mu_t$ allows for temporal trends and $\varepsilon_{it}$ are the zero-mean error terms which are independent of $d_{js},\boldsymbol{x}_{js},\kappa_{j},\mu_s$ for all $i,j,t,s$. 
Lagged outcomes and/or transformations of $\boldsymbol{x}_{it}$ can be included as extra covariates in the linear DID model \citep{Jones2011}. 
The parameters of the linear DID model can be estimated by ordinary least squares (OLS) regression (see \citet[pp. 167]{Angrist2009} for details). 
Let $\hat{\tau}_{it}^{\mathrm{DID}}$ denote the resulting estimate of $\tau_{it}$.

\textcolor{black}{
The linear DID model \eqref{eq:didb} makes very strong assumptions regarding the data generating mechanism. 
The term $\kappa_i$ in \eqref{eq:didb} allows expected counterfactual treatment-free outcomes to be higher (or) lower in the treated units than in control units, even after adjusting for observed covariates $\boldsymbol{x}_{it}$. 
Hence, $\kappa_i$ can represent an unobserved confounder. 
However, Equation \eqref{eq:didb} assumes that the effect of this possible confounder on the outcome is constant over time. 
Similarly, the term $\mu_t$ in \eqref{eq:didb} can only account for temporal trends that are common to both treated and control units. 
}

\textcolor{black}{
Although the linear DID specification \eqref{eq:didb} is often preferred  in practice due to its simple interpretation and implementation, there exist other methods that build on the parallel trends idea. 
\citet{Athey2006} relax the linearity assumption of \eqref{eq:didb}, allowing the outcome $y_{it}$ to be a more general (non-linear) function of the unobserved characteristics of unit $i$. 
However, it is difficult to implement their method when there are more than two time points and $\boldsymbol{x}_{it}$ is high-dimensional. 
Another method based on parallel trends is the triple differences method \citep{Atanasov2016,Wing2018}, which uses two groups of control units. 
For example, when the treated group consists of male employees of a company, then the control group can be either the female employees of the same company, or the male employees of a different company. 
In such situations, the triple differences method can use the second control group to correct for biases caused by the violation of the assumption of parallel trends between the outcomes of the treated group and the control group. 
}

There are many examples of the use of DID models. 
\citet{Ashenfelter1978} and \cite{Ashenfelter1985} investigate the effect of training programs on worker earnings. 
\citet{Card1990} assesses the impact that the Mariel Boatlift, a mass migration of Cuban citizens to Miami in 1980, had on the city's labour market, using four other cities as controls. 
\citet{Card1994} estimate the effect that the increase of the minimum salary had on employment rates in New Jersey's fast-food industry in 1992, using fast-food restaurants located in Pennsylvania as the control group. 
See \citet{Galiani2005,Branas2011,King2013} for recent works applications of the DID approach.

 \subsection{Latent factor models}\label{sec:lfm}
 \textcolor{black}{
 In the linear DID model of Equation \eqref{eq:didb}, there is one unit-specific term, $\kappa_i$, and this can represent a single unobserved confounder whose effect on the outcome is constant over time. 
 In the following latent factor model (LFM), $\kappa_i$ is replaced by $\boldsymbol{\lambda}_i\boldsymbol{f}_t$
 \begin{eqnarray}\label{eq:factor}\nonumber
  \quad y_{it}&=&y_{it}^{(0)}+\tau_{it} d_{it} \\
   y_{it}^{(0)}&=&\boldsymbol{x}^\top_{it}\boldsymbol\theta + \boldsymbol\lambda_i^\top\boldsymbol{f}_t +\varepsilon_{it},
  \end{eqnarray} 
where $\boldsymbol{f}_t=(f_{1t},\dots,f_{Jt})^\top$ are $J$ time-varying factors, $\boldsymbol\lambda_i=(\lambda_{i1},\dots,\lambda_{iJ})^\top$ are unit-specific factor loadings, and $\varepsilon_{it}$ are the zero-mean errors which are independent of $d_{js},\boldsymbol{x}_{js},\boldsymbol\lambda_{j},\boldsymbol{f}_s$ for all $i,j,t,s$. 
 When $\boldsymbol{f}_t=(1,\mu_t)^\top$ and $\boldsymbol\lambda_i=(\kappa_i,1)^\top$, the second line in \eqref{eq:factor} reduces to the second line in \eqref{eq:didb}. 
 So, the linear DID model is a special case of the LFM. 
 Just as $\kappa_i$ in the linear DID model can represent a single unobserved confounder, $\boldsymbol\lambda_i$ can represent $J$ unobserved confounders, whose effect on the outcome varies with time and is described by $\boldsymbol{f}_t$. 
 Hence, the LFM \eqref{eq:factor} relaxes the DID assumption that the average outcomes of control and treated units follow parallel trends. 
 In econometrics, $\boldsymbol{f}_t$ is interpreted as a `shock' that affects all units at time $t$ and $\boldsymbol\lambda_i$ represents the response of unit $i$ to these shocks \citep{Bai2009}. 
 }

\citet{Xu2017} proposes a three-step estimation procedure for predicting counterfactual treatment-free outcomes using the LFM model. 
In the first step, observations on control units are used to estimate $\boldsymbol\theta$, $\boldsymbol{f}_1,\ldots,\boldsymbol{f}_T$ and $\boldsymbol\lambda_1,\ldots,\boldsymbol\lambda_{n_1}$  through the iterative procedure of \citet{Bai2009} that minimises $\sum_{i=1}^{n_1}\sum_{t=1}^{T}(y_{it}-\hat{y}_{it}^{(0)})^2$, the mean squared error (MSE) between the observations $y_{it}$ and the corresponding predicted values $\hat{y}_{it}^{(0)}=\boldsymbol{x}_{it}^\top\hat{\boldsymbol\theta}+\hat{\boldsymbol\lambda}^\top_i\hat{\boldsymbol{f}}_t$. 
In the second step, the estimated factor loadings for the treated units, $\hat{\boldsymbol\lambda}_i$ ($i>n_1$), are obtained conditional on the parameter estimates obtained in the first step by minimising the MSE between $y_{it}$ and $\hat{y}_{it}^{(0)}$ for the treated units in the pre-intervention period. 
Finally, the third step involves estimating the intervention effects $\tau_{it}$ as $\hat\tau_{it}^{\mathrm{XU}}=y_{it}-\hat{y}_{it}^{(0)}$.

\textcolor{black}{
Several authors have proposed alternative methods for predicting counterfactuals using the LFM. 
These include \citet{Ahn2013,Gobillon2016,Chan2016} and \citet{Athey2017}. 
We have focused on the method of \citeauthor{Xu2017} because, to the best of our knowledge, it is the only one for which an R package has been developed.
}

\citet{Gobillon2016} and \citet{Xu2017} describe applications of the LFM to real data. 
\citet{Gobillon2016} estimate the effect on unemployment rates of a French program offering tax reliefs to companies that hired at least 20\% of their personnel from the local labour force. 
\textcolor{black}{\citet{Xu2017} evaluates the impact of Election Day Registration (EDR), a law that enables eligible citizens to register on site when they arrive at the voting centre, on voter turnout in the US}. 
For more applications of the LFM see \citet{Kim2014} and \citet{Navarro2018}.

\subsection{\textcolor{black}{Synthetic control-type approaches}} \label{sec:sc}
The original synthetic controls method (SCM) was developed by \citet{Abadie2003a} and \citet{Abadie2010} and can only be applied to one treated unit at a time. 
The idea behind the SCM is to find weights $\boldsymbol{w}=\left(w_1,\dots,w_{n_1}\right)^\top$ for the control units such that the weighted average of the controls' outcomes best predicts (in terms of MSE) the outcome of the treated unit during the pre-intervention period, and then use the weights to estimate the counterfactual treatment-free outcomes in the post-intervention period.
The set of weights $\boldsymbol{w}$ minimises
\begin{equation} \label{eq:synth1}
\sqrt{\left(\boldsymbol{y}_{n_1+1,1:T_1}-\boldsymbol{Y}_{1:T_1}^{\mathrm{c}}\boldsymbol{w}\right)^\top\boldsymbol{V}\left(\boldsymbol{y}_{n_1+1,1:T_1}-\boldsymbol{Y}_{1:T_1}^\mathrm{c}\boldsymbol{w}\right)},
\end{equation}
subject to the constraints 
\begin{equation}\label{eq:scw}
\sum_{i=1}^{n_1}{w_i}=1 \mbox{ and } w_i\geq0,
\end{equation}
where $\boldsymbol{Y}_{1:T_1}^\mathrm{c}$ is the $T_1\times n_1$ matrix with $i$-th column $\boldsymbol{y}_{i,1:T_1}$ and $\boldsymbol{V}$ is a $T_1\times T_1$ symmetric, positive semi-definite matrix reflecting the importance given to the different pre-intervention time points \citep{Abadie2003a}. 
The predicted counterfactual of the treated unit is:
\begin{equation}\label{eq:synth2}
\hat{\boldsymbol{y}}^{(0)}_{n_1+1,(T_1+1):T}=\boldsymbol{Y}_{(T_1+1):T}^{\mathrm{c}}\boldsymbol{w},
\end{equation}
where $\boldsymbol{Y}_{(T_1+1):T}^{\mathrm{c}}$ is defined analogously to $\boldsymbol{Y}_{1:T_1}^{\mathrm{c}}$. 
The estimated intervention effect at times $t$ ($t>T_1$) is then $\hat{\tau}_{n_1+1,t}^\mathrm{SCM}={y}_{n_1+1,t}-\hat{{y}}^{(0)}_{n_1+1,t}$.

It is also possible to use the covariates, by replacing $\boldsymbol{y}_{i,1:T_1}$ with $\boldsymbol{z}_{i,1:T_1}=(\boldsymbol{y}_{i,1:T_1}^\top,\boldsymbol{x}_{i,1:T_1}^\top)^\top$ in Equation \eqref{eq:synth1}. 
\citet{Abadie2010} suggest that instead of using the full data $\boldsymbol{z}_{i,1:T_1}$, it may be reasonable to consider only a few summaries, such as the mean outcome $\frac{1}{T_1}\sum_{t=1}^{T_1}{y_{it}}$ in the pre-intervention period, and the corresponding means of the covariates \textit{i.e}.\ to replace $\boldsymbol{z}_{i,1:T_1}$ by $(\frac{1}{T_1}\sum_{t=1}^{T_1}{y_{it}},\frac{1}{T_1}\sum_{t=1}^{T_1}{\boldsymbol{x}^\top_{it}})^\top$. 
Such reduction of the dimensionality of $\boldsymbol{z}_{i,1:T_1}$ might be necessary in applications with $T_1>>n_1$ in order to reduce computation time. 

The choice of matrix $\boldsymbol{V}$ can either be based on a subjective judgement of the relative importance of the variables in $\boldsymbol{y}_{i,1:T_1}$ or $\boldsymbol{z}_{i,1:T_1}$ or be determined through a data-driven approach. 
For example, \citet{Abadie2003a} and \citet{Abadie2010} choose $\boldsymbol{V}$ as the positive definite diagonal matrix that minimises the MSE between the observed outcomes $\boldsymbol{y}_{i,1:T_1}$ ($i=1,\dots,T_1$) and estimated outcomes $\hat{\boldsymbol{y}}_{i,1:T_1}=\boldsymbol{Y}_{1:T_1}^{\mathrm{c}}\boldsymbol{w}(\boldsymbol{V})$ in the pre-intervention period, where $\boldsymbol{w}(\boldsymbol{V})$ is the solution to \eqref{eq:synth1} for a fixed $\boldsymbol{V}$.

The SCM makes no assumptions regarding the data generating mechanism. 
The method has strong links with the matching literature, where the outcome of each treated individual is compared to the outcomes of controls with similar covariate values \citep{Rosenbaum2002,Stuart2010}. 
However, it is more general in the sense that a good match is sought by weighted averaging of the controls. 
The SCM also relates to the method of analogues used for time-series prediction. 
The difference is that in the method of analogues there is only one time-series and `controls' are simply earlier segments of the time-series; for more details see \citet{Viboud2003}.

There have been several proposed extensions of the SCM. 
To allow for multiple treated units, \citet{Kreif2016} apply the SCM to the averaged vector outcome $\bar{\boldsymbol{y}}^\mathrm{tr}=(\frac{1}{n_2}\sum_{i=n_1+1}^{n}{y_{i1}},\dots, \frac{1}{n_2}\sum_{i=n_1+1}^{n}{y_{iT}})^\top$ of the treated units. 
\citet{Acemoglu2016} assume that the intervention effects $\boldsymbol{\tau}_{n_1+1},\dots,\boldsymbol{\tau}_{n}$ are equal and estimate the common effect at time $t$ ($t>T_1$) as the weighted average $\sum_{i=n_1+1}^{n}{q_i^{-1}\hat{\tau}_{it}}/
\sum_{i=n_1+1}^{n}{q_i^{-1}}$, where $\hat{\tau}_{it}$ ($i>n_1$) is obtained by applying the original algorithm to just the data on treated unit $i$ and the control units $1,\ldots,n_1$, and $q_i=\sqrt{T_1^{-1}(\boldsymbol{y}_{i,1:T_1}-\hat{\boldsymbol{y}}^{(0)}_{i,1:T_1})^\top(\boldsymbol{y}_{i,1:T_1}-\hat{\boldsymbol{y}}^{(0)}_{i,1:T_1})}$. 
Their stated rationale for using weights $q_i^{-1}$ is that units with good fit in the pre-intervention period should be more reliable for estimating the common intervention effect and hence receive higher weights. 
 
\textcolor{black}{\citet[henceforth HCW]{Hsiao2012} and \citet[henceforth DI]{Imbens2016} extend the SCM by adding a time-constant intercept term to the SCM estimator and removing the constraints on the weights. 
The intercept is necessary when the outcome of the treated unit is systematically (over time) higher or lower than the outcomes of the controls units and hence there exists no set of weights that can provide a good fit for $y_{n_1+1,t}$ in the pre-intervention period. 
The removal of the constraints on the weights is useful, for example, when there exist control units with outcomes that are negatively correlated with the outcomes on the treated unit. 
HCW suggest estimating $y_{n_1+1,t}^{(0)}$ ($t>T_1$) as $\hat{y}_{n_1+1,t}^{(0)}=\beta_0+\sum_{i=1}^{n_1}{\beta_iy_{it}}$, where $\beta_0,\dots,\beta_{n_1}$ are the OLS coefficient estimates of the regression of $\boldsymbol{y}_{n_1+1,1:T_1}$ on $\boldsymbol{y}_{1,1:T_1},\dots,\boldsymbol{y}_{n_1,1:T_1}$, \textit{i.e.} they minimise \begin{equation}\label{eq:net}
\left(\boldsymbol{y}_{n_1+1,1:T_1}-\beta_0\boldsymbol{1}-\boldsymbol{Y}_{1:T_1}^{\mathrm{c}}\boldsymbol{\beta}\right)^\top\left(\boldsymbol{y}_{n_1+1,1:T_1}-\beta_0\boldsymbol{1}-\boldsymbol{Y}_{1:T_1}^{\mathrm{c}}\boldsymbol{\beta}\right),
\end{equation}
where $\boldsymbol{1}$ denotes a $T_1$-vector of ones, $\beta_0$ is the intercept and $\boldsymbol{\beta}=\left(\beta_1,\ldots,\beta_{n_1}\right)^\top$. 
\citet{Amjad2017} also remove the constraint on the weights and suggest that, before estimating these weights, the data on the control outcomes $\boldsymbol{Y}_{1:T_1}^{\mathrm{c}}$ should be de-noised.
}

\textcolor{black}{
\citet{Ben2018} introduced the augmented SCM. 
First, the SCM is applied and weights $w_1,\ldots,w_{n_1}$ obtained. 
Second, a model (e.g.\ a LFM) for the untreated outcomes $y_{i}^{(0)}$ of all $n_1+1$ units is fitted to all the outcomes of the untreated units and the pre-intervention outcomes of the untreated unit. 
If $\tilde{y}^{(0)}_{it}$ ($i=1,\ldots, n_1+1; t>T_1$) denote the predicted untreated outcomes from this model, then $\sum_{i=1}^{n_1} w_iy_{it} - \tilde{y}^{(0)}_{n_1+1,t}$ is an estimate of the bias of the
SCM estimator. 
The augmented SCM estimator of the counterfactual $y_{n_1+1}^{(0)}$ equals the original SCM estimate plus this estimated bias. 
They argue that this method is particularly useful when the SCM method provides a poor fit in the pre-intervention period. 
}

\textcolor{black}{ 
\citet{Hazlett2018} estimate the weights using a kernel transformation of the pre-intervention outcomes. 
This is done to ensure that higher-order features of the outcomes (authors mention, \textit{e.g.}, volatility and variance) are taken into account when estimating the weights. 
Using simulated examples, they showed that their approach can eliminate biases that occur if the untransformed outcomes are used to estimate the weights.
}

Several recent works utilise synthetic control-type approaches for estimating the effects of an intervention.
These include \citet{Cavallo2013}, who examine the effect of large-scale natural disasters on gross domestic product, and \citet{Ryan2016}, who investigate the impact that UK's Quality and Outcomes Framework, a pay-for-performance scheme in primary health, had on population mortality. 
For more applications of synthetic control-type methods, see \citet{Billmeier2013,Fujiki2015,Saunders2015} and \citet{Aytuug2017}.

\subsection{Causal impact}\label{sec:cim}
The causal impact method (CIM) was introduced by \citet{Brodersen2015} and can only be applied to a single treated unit at a time. 
A Bayesian model is assumed for the outcome of the treated unit. 
This model includes a time-series component that relates the outcome of the treated unit at time $t$ to previous outcomes on the same unit, and a regression component that uses the outcomes on control units as covariates. 
Specifically:
\begin{eqnarray}\label{eq:bstseg}\nonumber
y_{n_1+1,t}^{(0)} &=& \beta_{0t}+\sum_{i=1}^{n_1}\beta_iy_{it}+\varepsilon_{t} \hspace{1cm} (t=1,\dots,T)
\\ \nonumber
\beta_{0,t+1} &=& \beta_{0,t}+\delta_{t}+\eta_{t}\\  
\delta_{t+1} &=& \delta_{t}+\zeta_{t},
\end{eqnarray}
with mutually independent $\varepsilon_{it}\sim\mathrm{N}(0,\sigma_\varepsilon^2)$, $\eta_t\sim\mathrm{N}(0,\sigma_\eta^2)$ and $\zeta_t\sim\mathrm{N}(0,\sigma_\zeta^2)$, and priors for $\beta_{00}$, $\delta_0$, $\beta_1,\ldots,\beta_{n_1}$, $\sigma_\varepsilon^2$, $\sigma_\eta^2$ and $\sigma_\zeta^2$. 
In Equations \eqref{eq:bstseg}, the component $\beta_{0t}$ induces temporal correlation in the outcome, the regression component $\sum_{i=1}^{n_1}\beta_iy_{it}$ relates $y_{n_1+1,t}^{(0)}$ to measurements from control units, and the error component $\varepsilon_{t}$ accounts for unexplained variability. 
More complex models can be adopted \citep{Brodersen2015}, \textit{e.g}.\ by adding a seasonal component. 

The model \eqref{eq:bstseg} is fitted to the observed data, $y_{n_1+1,1}^{(0)}, \ldots, y_{n_1+1,T_1}^{(0)}$, treating the counterfactuals $y_{n_1+1,T_1+1}^{(0)}, \ldots, y_{n_1+1,T}^{(0)}$ as unobserved random variables. 
Independent, improper, uniform priors are used for $\tau_{n_1+1,T_1+1}, \ldots,\tau_{n_1+1,T}$. 
Then, $L$ samples $y_{n_1+1, t}^{(0, l)}$ $(l=1, \ldots, L$) are drawn from the resulting posterior predictive distribution of the counterfactual outcome $y_{n_1+1, t}^{(0)}$ ($t > T_1$), thus providing samples $y_{n_1+1,t}-y_{n_1+1, t}^{(0, l)}$ from the posterior distribution of $\tau_{n_1+1,t}$.
Typically, this would be done using a Markov chain Monte Carlo algorithm. 
A point estimate $\hat{\tau}_{n_1+1,t}^\mathrm{CIM}$ for the causal effect $\tau_{n_1+1,t}$ at time $t$ ($t>T_1$) is then given by its posterior mean.

\citet{Bruhn2017} use the CIM to assess the impact of pneumococcal conjugate vaccines on pneumonia-related hospitalisations using hospitalisations from other diseases as the control time-series. 
\citet{Vocht2017} evaluate the benefits of stricter alcohol licensing policies on alcohol-related hospitalisations in several areas, control areas being other areas where these policies were not implemented. 
See also \citet{DeVocht2016,Gonzalez2016,Vizzotti2016} for other applications of the CIM.

%% file: comparison.tex
\section{Quantification of uncertainty \& hypothesis testing}\label{sec:uncertainty}
\textcolor{black}{We now describe approaches to estimating standard errors and testing the null hypothesis that $\tau_{it}=0$, or, for Bayesian methods, estimating the posterior distribution of $\tau_{it}$.}

\textbf{DID}. 
\textcolor{black}{If it is assumed that the errors $\varepsilon_{it}$ in the linear DID are mutually independent and homoscedastic, variance estimates for the OLS estimates of $\tau_{it}$ ($i>n_1,t>T_1$) are easy to obtain. 
These represent the variance of $\hat{\tau}_{it}^{\mathrm{DID}}$ over repeated samples of the errors $\varepsilon_{it}$ holding $(\boldsymbol{x}_{11}^\top,\ldots,\boldsymbol{x}_{1T}^\top,\kappa_1,\boldsymbol{d}_1^\top,\ldots,\boldsymbol{x}_{n1}^\top,\ldots,\boldsymbol{x}_{nT}^\top,\kappa_n,\boldsymbol{d}_n^\top,\\\mu_1,\ldots,\mu_T)^\top$ fixed. 
A Wald test for $\tau_{it}=0$ can then be performed.
} 
However, the assumption that the errors $\varepsilon_{it}$ are mutually independent may not be plausible. 
\citet{Bertrand2004} show that when, as is likely, the errors $\varepsilon_{i1},\ldots,\varepsilon_{iT}$ are serially correlated, the variance estimator for $\hat{\tau}_{it}$ is biased downwards and type-I error rates are inflated, and they describe methods to deal with this. 
Standard errors can also be underestimated if there are correlations due to units being grouped (e.g. hospitals within the same county); \citet{Donald2007} discuss possible solutions.

\textbf{LFM}. \citet{Xu2017} uses parametric bootstrap to obtain confidence intervals for $\hat\tau_{it}^\mathrm{XU}$ and $p$-values, assuming that $\varepsilon_{1t},\ldots,\varepsilon_{nt}$ are independent and homoscedastic at each individual time $t$. 
Repeated sampling here is of the errors $\varepsilon_{it}$ holding $(\boldsymbol{x}_{11}^\top,\ldots,\boldsymbol{x}_{1T}^\top,\boldsymbol\lambda_1^\top,\boldsymbol{d}_1^\top,\ldots,\boldsymbol{x}_{n1}^\top,\ldots,\boldsymbol{x}_{nT}^\top,\boldsymbol\lambda_n^\top,\boldsymbol{d}_n^\top,\boldsymbol{f}_1^\top,\ldots,\boldsymbol{f}_T^\top)^\top$ fixed.
\textcolor{black}{ 
\citet{Li2018} derive the asymptotic distribution (as $T_1,T_2\rightarrow\infty$)  of the average effect $\sum_{t=T_1+1}^{T}\hat{\tau}_{it}^\mathrm{XU}$ for the $i$-th treated unit.
}

\textcolor{black}{
\textbf{Synthetic control-type approaches.} 
\citet{Abadie2010,Abadie2015} argue that
traditional statistical inference is difficult in this setting, unless one is prepared to assume that the unit that received the intervention was chosen at random. 
Under that assumption, a standard permutation test would provide a valid $p$-value for the null hypothesis that treatment would have no effect on any of the units (\textit{i.e.} $\tau_{it}=0$ for all $i=1,\ldots,n_1+1$). 
\citeauthor{Abadie2015} propose using a very similar test even in settings where the intervention is not randomly assigned and called this a `placebo test'. 
They argue that such a test provides an alternative mode of inference, saying that our confidence that a large treatment effect estimate truly reflects the effect of the intervention would be undermined if similarly large effect estimates were obtained when the treatment labels of the units were permuted.
}

More specifically, \citet{Abadie2010,Abadie2015} compare $\hat{\tau}_{n_1+1,t}^\mathrm{SCM}$ to $\hat{\tau}_{1t}^\mathrm{SCM},\dots,\hat{\tau}_{n_1t}^\mathrm{SCM}$, the estimated effects considering each of the control units in turn as though it had been the treated unit, and using the remaining $n_1-1$ controls to estimate the weights, at each post-intervention time. 
Their test statistic $r_i$ is 
\begin{equation}
\label{eq:sctest}
r_{i}=
\frac{T_1( \boldsymbol{y}_{i,T_1+1:T}-\hat{\boldsymbol{y}}_{i,T_1+1:T}   )^\top( \boldsymbol{y}_{i,T_1+1:T}-\hat{\boldsymbol{y}}_{i,T_1+1:T}   ) }{T_2( \boldsymbol{y}_{i,1:T_1}-\hat{\boldsymbol{y}}_{i,1:T_1}   )^\top( \boldsymbol{y}_{i,1:T_1}-\hat{\boldsymbol{y}}_{i,1:T_1}   )}
,  
\end{equation}   
that is, the ratio of post- to pre-intervention MSE between the observed and predicted outcomes. 
The predicted counterfactual of control unit $i$ ($i\leq n_1$) is obtained by applying the SC method to that unit, using the remaining $n_1-1$ controls to find weights. 
Their intuition is that under the null hypothesis the predictive ability of the SCM should be similar in the two periods and thus the ratio $r_{n_1+1}$ close to 1. 
Hence, a value of $r_{n_1+1}$ that lies in the tail of the empirical distribution of $r_1,\dots,r_{n_1+1}$ can be viewed as evidence for a non-zero intervention effect. 
\textcolor{black}{\citet{Firpo2018} investigate the impact that the choice of the test statistic has on the results of \citeauthor{Abadie2015}'s test, finding that $r_i$ outperformed alternative statistics that they considered in several performance measures.}

\textcolor{black}{
\citet{Firpo2018} propose a generalisation of \citeauthor{Abadie2015}'s placebo test. 
Rather than giving equal weight to all possible permutations of the treatment labels when calculating the $p$-value, they make the weights depend on a sensitivity parameter $\phi$. 
The reasoning is that even if the unit that received the intervention had actually been chosen at random, some units might have been more likely to be chosen, thus making some permutations of the treatment labels more probable than others. 
\citet{Firpo2018} vary the value of $\phi$ and assess how robust to this value is the conclusion of a treatment effect (or lack thereof).
}

\citet{Amjad2017} take an empirical Bayes approach to test the hypothesis that $\tau_{n_1+1,t}=0$. 
They assume that $\boldsymbol{y}_{n_1+1,1:T}\sim\mathcal{N}\left(\tilde{\boldsymbol{Y}}_{1:T_1}^{\mathrm{c}}\boldsymbol{w},\sigma^2\boldsymbol{I}\right)$, where $\tilde{\boldsymbol{Y}}_{1:T_1}^{\mathrm{c}}$ are the de-noised control outcomes obtained via singular value thresholding, and the weights $\boldsymbol{w}$ have a $\mathcal{N}\left(0,\sigma^2_{\boldsymbol{w}}\boldsymbol{I}\right)$ prior distribution, for some value of $\sigma^2_{\boldsymbol{w}}$. 
The posterior distribution of $\boldsymbol{w}$ can be used to calculate the posterior predictive distribution of $y_{n_1+1,t}^{(0)}$ ($t>T_1$). 
Let $a$ and $b$ be the  97.5th and 2.5th centiles of this distribution. 
The 95\% posterior credible interval for $\tau_{n_1+1,t}$ is $(y_{n_1+1,t} - a, y_{n_1+1,t} - b)$. 

\textcolor{black}{
\citet{Abadie2010} also consider a variant of their placebo test in which the time of the intervention, rather than the unit that receives intervention, is changed. 
They do not, however, propose this as being a way to calculate a $p$-value.
}

\textcolor{black}{ 
In their application, HCW fit an autoregressive model to the estimated intervention effects $\hat{\tau}_{n_1+1,T_1+1}^\mathrm{HCW},\ldots,\hat{\tau}_{n_1+1,T}^\mathrm{HCW}$. 
They then test the null hypothesis that the mean of these effects, which they refer to as the long-run intervention effect, equals zero. 
In their implementation of the HCW method, \citet{Gardea2017} use a test that is equivalent to the test proposed by \citet{Abadie2010,Abadie2015} for the SCM, to test if $\tau_{it}=0$ for all $i$ and $t>T_1$. 
As pointed out by one of the referees, an intuitive approach to obtain confidence intervals for $\hat{\tau}_{n_1+1,t}^\mathrm{HCW}$ would be to use the bootstrap. 
Finally, \citet{Li2017} derive the asymptotic distribution (as $T_1,T_2\rightarrow\infty$)  of the average effect $\sum_{t=T_1+1}^{T}\hat{\tau}_{n_1+1,t}^\mathrm{HCW}$.   
}

\textbf{CIM}. For the CIM, a 95\% posterior credible interval for $\tau_{n_1+1,t}$ can be calculated as $(y_{n_1+1,t} - a, y_{n_1+1,t} - b)$, where $a$ and $b$ denote, respectively, the 97.5th and 2.5th centiles of the posterior predictive distribution of the counterfactual $y_{n_1+1, t}^{(0)}$.

 \textcolor{black}{ 
\textbf{All methods}. Recently, there has been work building upon the end-of-sample stability test \citep{Andrews2003}. 
For a single treated unit, the idea is that under the hypothesis of no intervention effect, the process $y_{n_1+1,t}-\hat{y}_{n_1+1,t}^{(0)}$ ($t=1,\ldots,T$) is stationary. 
\citet{Cherno2017} propose a permutation procedure to test the stationarity of this process and show that their approach gives valid inference for several methods including DID, LFM and SCM. 
\citet{Hahn2017} apply the same idea to the SCM. 
Both note that confidence sets for the intervention effect can be obtained by statistic inversion. 
}

\section{Implementation issues}\label{sec:practice}
In this section, we discuss issues related to the practical implementation of the methods presented in Section \ref{sec:met}: model choice and diagnostic checks.

\subsection{Model choice}\label{sec:model}

\textit{Choosing the control units}. 
When implementing the SCM, HCW, DI and CIM, it may be desirable to exclude some of the potential control units. 
Using all potential controls might result in non-unique causal effect estimates when there are more such controls than pre-intervention time points. 
Moreover, standard errors of estimates can be reduced by discarding controls whose outcomes are not related to the outcome of the treated unit.

HCW develop a two-stage approach to exclude potential controls.
For each $\ell=1,\dots,n_1$ they implement their method ${n_1 \choose \ell}$ times, where each time they use a different subset of size $\ell$ of the control units. 
For each $\ell$ they choose the subset that maximises the regression $R^2$ and thus obtain $n_1$ candidate models. 
They recommend choosing one of these $n_1$ models according to a model selection criterion such as the AIC.
An alternative approach was suggested by \citet{Li2017}, who use the least absolute shrinkage and selection operator (LASSO) to select controls.

DI exclude potential controls by encouraging some of the weights $\beta_i$ to shrink towards (or even equal to) zero. 
This achieved by including the penalty term
\begin{equation}\label{eq:scpen}
\rho\left(\frac{1-\delta}{2}\sum_{i=1}^{n_1}\beta_i^2+\phi\sum_{i=1}^{n_1}|\beta_i|\right), 
\end{equation}
in the objective function \eqref{eq:net}, where $\rho$ and $\phi$ are penalty parameters. 
For CIM \citet{Brodersen2015} induce sparsity on the vector $(\beta_1,\ldots,\beta_{n_1})^\top$ that describes the dependence on controls by using a spike-and-slab prior.

\textit{Choosing the covariates}. 
An issue that may arise when implementing the linear DID and the LFM methodology of \citet{Xu2017} is the choice of covariates to include. 
Exclusion of potential covariates may be desirable for the same reasons that one might exclude control units. 
For the linear DID model, covariates, which may include lagged outcomes and interactions of lagged outcomes with the covariates, may be selected by imposing sparsity on the regression coefficient vector, using, for example, the LASSO. 
\textcolor{black}{For the LFM, one can use the factor-lasso approach of \citet{Hansen2016}.}

When implementing the SCM, users need to decide which variables (pre-intervention outcomes, covariates or summaries of these) to use to determine the weights. 
\citet{Ferman2016b} demonstrate that the estimated counterfactual may differ depending on which variables are used. 
\citet{Dube2015} develop the following approach for selecting among $K$ sets of variables. 
First, for every set $k$ ($k=1,\dots,K$) they apply the SCM to the data on every control unit in turn, and calculate the predicted outcomes $\hat{y}_{it}^{(k)}$ ($i=1,\dots,n_1$) based on the estimated weights. 
Then, they choose the set $k^\ast$ that minimises the mean (over control units) MSE between observed and predicted outcomes $\hat{y}_{it}^{(k)}$ in the post-intervention period. 
An alternative approach when $T_1$ is large is to split the pre-intervention data into a training dataset, to which the SCM is applied using different sets of variables, and a validation dataset, which is used to assess which set has the best predictive performance.

\textit{Other issues}. 
Some of the methods for estimating the parameters of the LFM \citep{Gobillon2016,Chan2016,Xu2017} require that the number of factors $J$ be chosen. 
The usual approach is to fit the LFM for various values of $J$ and determine the optimal $J$ using cross-validation. 
An alternative for choosing $J$ is to use the procedures developed by \citet{Bai2002}. 
However, these approaches provide estimates of standard errors that do not account for the uncertainty about $J$.

For the CIM, practitioners need to decide what dynamical components to include in the counterfactual model. 
Similar methods to those used for choosing variables for the SCM can be used. 
An alternative is to fit several models and use the one that achieves the optimal trade-off between accuracy (the difference $y_{n_1+1,t}-\hat{y}_{n_1+1,t}$) and precision (the length of the credible interval for $y_{n_1+1,t}$) in the pre-intervention period. 
In small datasets the inferences provided by the CIM impact method will be sensitive to the choice of prior distributions. 
Therefore, these specifications should ideally be determined based on expert opinion.

\subsection{Diagnostics} \label{sec:blacks}
All the methods described in this article make assumptions about the counterfactual outcomes of the treated units in the post-intervention period. 
Since these outcomes cannot be observed, it is never possible to test the full set of assumptions. 
Nonetheless, it is sometimes possible to assess the validity of a subset of these assumptions using data from the pre-intervention period. 

When no covariates are used and $T_1>1$, an informal check of the parallel trends assumption of DID methods can be conducted by plotting the average outcomes of control and treated units in the pre-intervention period \citep{Keele2013}: an approximately constant (over time) distance between the two lines suggests that parallel trends is plausible. 
The SCM should not be used when the outcome of the treated unit lies outside the convex hull of the outcomes of controls units. 
This can be checked by plotting the time-series of the outcome on all the units.

The fit provided for the outcomes on treated units in the pre-intervention period can be used as a diagnostic check. 
Intuitively, if a model is not predictive of the outcome in the pre-intervention period, it is less likely to provide good predictions for the counterfactuals in the post-intervention period. 
Goodness-of-fit can be assessed using the MSE between the observed and predicted values. 
However, in order for a good pre-intervention fit to be reassuring, one needs to establish that it does not occur due to overfitting, as can be the case \textit{e.g}. for the SCM when $n_1>T_1$. 
For the methods that provide fitted values for the outcomes of control units, \textit{i.e.}\ the linear DID model and the method of \citet{Xu2017}, one can further use the fit over the post-intervention period for these units as a diagnostic tool.

Finally, for both the linear DID model and the LFM method of \citet{Xu2017}, extrapolation biases may occur when the covariates (and loadings for the LFM) of treated and control units do not share a common support. 
In order to exclude the possibility of such biases, it suffices to ensure that the characteristics of the treated units are not extreme compared to the characteristics of control units. 
When a small number of covariates (and factors) is used, one can visually compare the two groups for each covariate (and loading) in turn. 
If this is not feasible, methods for multivariate outlier detection (\textit{e.g}.\ \citet{Filz2008}) can be used to identify treated units with extreme characteristics.

%% file: real.tex
\section{Application: Effect of German reunification on GDP} \label{sec:real1}
In this section, we demonstrate the use of the methods we have described by analysing the data introduced in Section \ref{sec:germany}. 
The dataset is publicly available\footnote{\url{http://dx.doi.org/10.7910/DVN/24714}}. 
\textcolor{black}{We omit the available covariates because they might have been affected by the reunification}. 

\textcolor{black}{
For the DID and SCM, some of the diagnostic checks described in Section \ref{sec:blacks} do not require implementing these methods and therefore we started by carrying out these tests. 
Figure \ref{fig:didpar} of Appendix \ref{sec:suppl} shows the difference between West Germany's GDP, $y_{17,t}$, and the average GDP in the control countries, $\frac{1}{16}\sum_{i=1}^{16}y_{it}$, over the pre-reunification period. 
The difference has a clear increasing trend suggesting that the parallel trends assumption does not hold, so the linear DID model is not appropriate for this application. 
As we see from Figure \ref{fig:germany}, the outcome of the treated unit lies in the convex hull of the outcomes of control units so this provides no evidence that the SCM should not be used.
}

We only implement methods for which (to the best of our knowledge) R \citep{R} software exists. 
The linear DID method can be implemented using any linear regression function (\textit{e.g}.\ \textit{lm}). 
For the remaining methods, we used the packages specifically developed for these methods: \textit{gsynth} for the LFM; \textit{Synth} \citep{Abadie2011} for the SCM; \textit{pamp} \citep{Vega2015} for the HCW method; and \textit{CausalImpact} for the CIM. 
The code we used for our real data analysis is available online\footnote{\url{https://osf.io/b5fv3/}}. 

\textcolor{black}{
We fitted the linear DID model \eqref{eq:didb}. 
For the method of \citet{Xu2017} we set $f_{1t}=1$ for all $t$ and $\lambda_{i2}=1$ for all $i$ in order to have time and country fixed effects, respectively. 
The total number of latent factors was set via cross-validation. 
For the SCM, we estimated the weights using the whole vector of outcomes $\boldsymbol{y}_{i,1:T_1}$ in the pre-intervention period (rather than summaries of the outcomes). 
The HCW method was implemented using all control countries and pre-intervention time points. 
Finally, for the CIM we fitted the model of Equation \eqref{eq:bstseg} but without the term $\delta_t$ because we found that inclusion of this term did not improve the fit and led to substantially wider credible intervals for the causal effect of interest. 
The prior distributions for all model parameters were set to the software defaults. 
We fitted the linear DID method for illustration purposes even though DID should not be used here. 
}     

\begin{figure}[htp]
        \centering
        \includegraphics[scale=1.0]{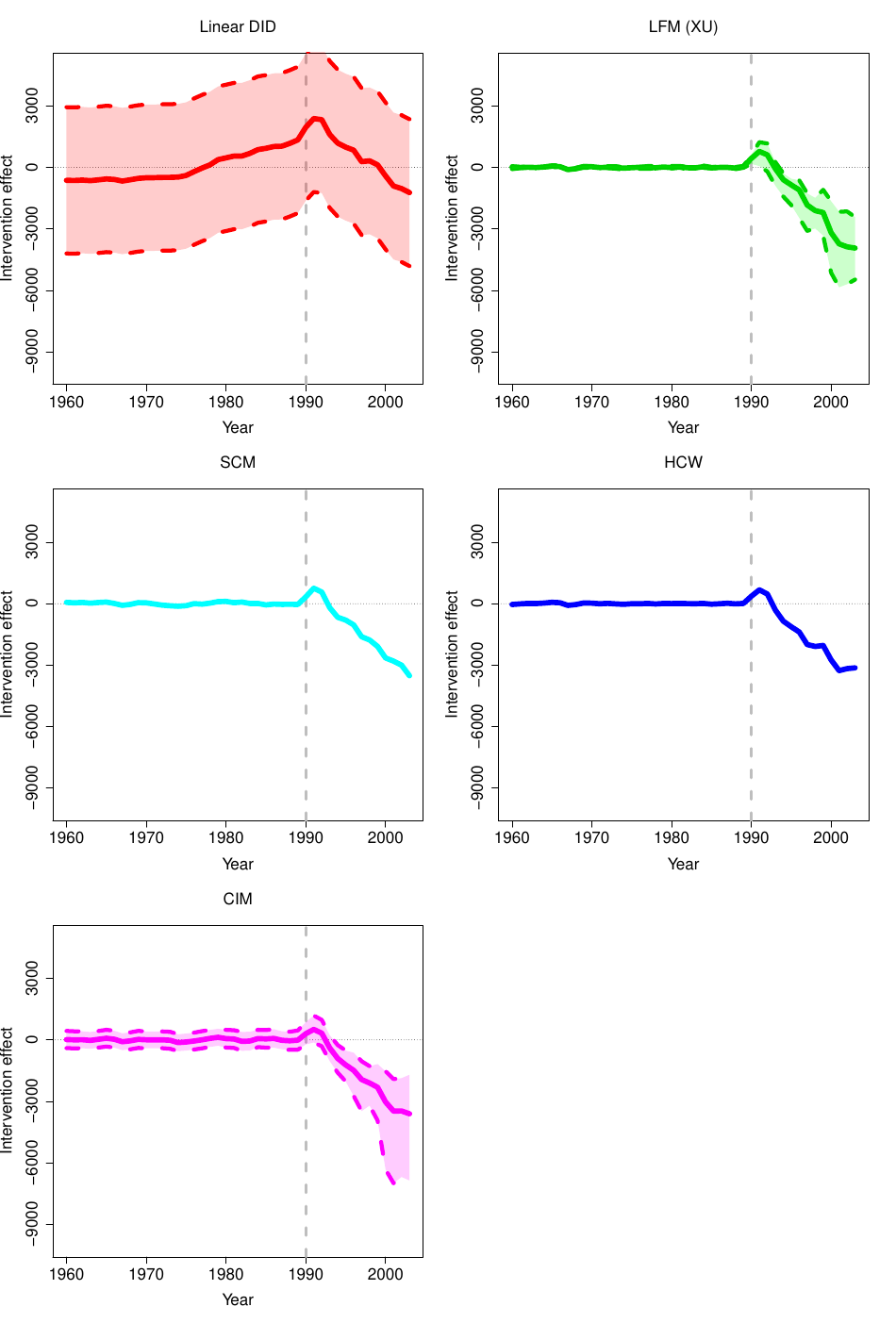}
        \vspace{-0.15in}
      \caption{Annual estimates of the effect of the German reunification on West Germany's per-capita GDP obtained using the linear DID model (red), the LFM (green), the SCM method (light blue), the HCW method (blue) and the CIM (purple). The dashed lines (when applicable) represent the 95\% confidence/credible intervals.}      
      \label{fig:eff}
 \end{figure}

\textcolor{black}{
Before examining the causal estimates, we performed the remaining diagnostic checks. 
Figure \ref{fig:eff} shows the difference between the actual and estimated counterfactual West German GDP, $\boldsymbol{y}_{17,\cdot}-\hat{\boldsymbol{y}}_{17,\cdot}$ for the entire study period. 
We see that all methods except for the linear DID almost perfectly reproduce West Germany's GDP before reunification. 
Thus, the pre-intervention goodness-of-fit diagnostic provides no indication against any of the methods except for linear DID.  
The estimated factor loadings for the 17 countries in the dataset are shown in Table \ref{tab:loadings} of Appendix \ref{sec:suppl}. 
The estimated loadings for West Germany are not extreme compared to the estimated loadings of the control countries, hence suggesting that the predicted counterfactual is not obtained by extrapolation. 
Overall we see that the only method that fails our diagnostic checks is the linear DID.}

\textcolor{black}{
Figure \ref{fig:eff} reveals that the other four methods provide similar estimates of the causal effect. 
In particular, the difference between the observed and counterfactual outcomes is positive during the first three years after 1989, suggesting that reunification initially had a positive impact on West Germany's GDP. 
\citet{Abadie2015} attribute this to a `demand boom'. 
The estimated impact reduces thereafter, and is negative for all four methods in year 2003. 
The estimated average reduction in annual GDP over the period 1990-2003 due to the reunification (we also show DID for completeness) is shown in Table \ref{tab:gdp}. 
}

\begin{table}[ht]
\begin{center}
\begin{tabular}{cr}
 \textbf{Method}&\textbf{GDP decrease} \\ \hline \hline
  Linear DID&   -604\\
  LFM (XU)&     1546\\
  SCM&           1322\\
  HCW&    1473\\
  CIM&           1629\\
\end{tabular}
\caption{Average (over the period 1990-2003) reduction in West Germany's annual per capita GDP, as estimated by the 5 methods. All values are in United States dollars.}
\label{tab:gdp}
\end{center}
\end{table}

Figure \ref{fig:eff} presents 95\% intervals for the LFM of \citet{Xu2017} and the CIM. 
These exclude zero in all years after 1993 thus suggesting a significant intervention effect. 
The placebo test of no intervention effect in any of the years 1990-2003 described by \citet{Abadie2010,Abadie2015} is also suggestive of a non-zero intervention effect. 
In particular, the $r$ statistic defined in Equation \eqref{eq:sctest} is $r_{17}=30.72$ for West Germany, larger than all the $r_i$ values obtained for the 16 control countries. 
We further implemented this test with the HCW method. 
The rank of the $r$ statistic for West Germany is 16, that is there is only one country whose $r$ statistic is higher. 
Table \ref{tab:rstat} in Appendix \ref{sec:suppl} shows the $r$ statistics obtained by applying the SCM and HCW methods. 
Overall, taking into consideration all tests conducted, we conclude that there is evidence that reunification had a negative long-term impact on West Germany's per-capita GDP, although it may have had a positive short-term impact.

%% file: discussion.tex
\section{Discussion}\label{sec:discussion} 

\subsection{Connections between methods}\label{sec:connections}
\textcolor{black}{
There are several ways in which the methods described in this paper relate to one another. 
} 

\textcolor{black}{
Firstly, for the case of a single treated unit and no covariates, most of them propose counterfactual estimators of the form $\hat{y}^{(0)}_{n_1+1,t}=\alpha_t+\sum_{i=1}^{n_1}\beta_iy_{it}$ ($t>T_1$) with $\alpha_t$ and $\beta_i$ being estimated using the data from the pre-intervention period (\textit{i.e.}\ $t\leq T_1$). 
For the DID method, the parallel trends assumption implies that $\alpha_t=\frac{1}{T_1}\sum_{s=1}^{T_1}\left(y_{n_1+1,s}-\frac{1}{n_1}\sum_{i=1}^{n_1}y_{is}\right)$ for all $t>T_1$ and $\beta_i=\frac{1}{n}$ for all $i\leq n_1$ \citep{Cherno2017}. 
The SCM assumes $\alpha_t=0$ for all $t$ and requires that $\beta_1,\ldots,\beta_n$ are non-negative and sum to one. 
The HCW and DI methods impose the constraint that  the intercept is constant over time, \textit{i.e.} $\alpha_t=\alpha$ for all $t$. 
Finally, the CIM assumes that $\alpha_t$ obeys a time-series model (\textit{e.g.}\ a random walk model). 
When there are covariates these similarities break down because methods account for covariates in a different way. 
}

\textcolor{black}{
Secondly, most of the methods relate to the LFM. 
We have already seen that the linear DID model \eqref{eq:didb} is a special case of the LFM \eqref{eq:factor}. 
As discussed in Appendix \ref{sec:theory}, the SCM and HCW estimators are asymptotically unbiased when the true data-generating mechanism obeys a LFM.
We expect that due to their similarities with the HCW estimator just explained, both DI and CIM estimators will be unbiased under the same LFM. 
}

\subsection{Recommendations for implementation}
\textcolor{black}{
None of the methods is universally superior to the others. 
Extensive simulation experiments comparing the relative performance of a subset of them have been conducted by multiple authors including \citet{Gobillon2016,ONeill2016,Gardea2017,Xu2017} and \citet{Kinn2018}. 
They all find settings in which one of the methods outperforms the others.  
However, the findings from these simulation studies may not generalise to other data generating mechanisms.  
Practitioners should choose the method to apply on the basis of the characteristics of the dataset and, in particular the values, of $n_1$, $T_1$ and their ratio $n_1/T_1$.}

\textcolor{black}{
The DID method can be used for any $n_1$ and $T_1$. 
As explained in Sections \ref{sec:did} and \ref{sec:connections}, the DID method arguably requires the strongest assumptions. 
As a result, it may provide more precise estimates of the intervention effects compared to the other methods. 
However, these estimates might be severely biased when the parallel trends assumption does not hold (see simulation studies by \citet{ONeill2016} and \citet{Gobillon2016}). 
This occurred in our application (Section \ref{sec:real1}), where the DID estimate of the average reunification effect had opposite sign compared to all the other estimates. 
Hence, it is essential to test the plausibility of parallel trends in the pre-intervention period before applying DID to a dataset. 
This is easy when there are no covariates. 
}

\textcolor{black}{
The LFM can be used for any value of $n_1/T_1$. 
However, both $n_1$ and $T_1$ should be at least moderate in size in order to accurately estimate the factors and loadings, respectively. 
For example, for the asymptotic unbiasedness property of \citeauthor{Xu2017}'s method (see Appendix \ref{sec:theory}) to be relevant to a finite sample, they recommend $T_1>10$ and $n_1>40$. 
}

\textcolor{black}{
Synthetic control approaches are mostly suited for applications where $T_1$ is large. 
This is required to accurately estimate the relationships between the outcome of the treated unit and the outcomes of control units. 
When $n_1\geq T_1$ regularisation is required because the number of parameters exceeds the number of observations\footnote{Synthetic control approaches regress the outcome of the treated unit on the outcomes of the control units. Therefore we can think of $\boldsymbol{y}_{1:n,t}$ as a single data point (observation) in a regression model.}. 
Regularisation is possible for both the HCW and DI estimators, as described in Section \ref{sec:model}, but not for the SCM. 
The SCM should not be used when the outcome of the treated unit does not lie in the convex hull of the outcomes of control units.  
}

\textcolor{black}{
The CIM is similar in spirit to synthetic control approaches and also requires large $T_1$. 
Because of its time-series component it can work even in cases when the outcome of treated unit is not correlated to the outcomes of control units. 
 However, it requires larger $T_1$ than synthetic control-type approaches to estimate the additional time-series parameters. 
In practice, the value of $T_1$ required will depend on the complexity of the time-series model. 
When $n_1>T_1$ regularisation can be achieved via a spike-and-slab prior on the regression coefficients.  
}

\textcolor{black}{
In applications where certain covariates are known to be highly predictive of the outcome, it is preferable to use the linear DID or LFM.
This is because they use the covariates of control units and therefore can estimate the regression parameters of the predictive covariates with higher precision compared to the HCW, DI and CIM\footnote{Although one might argue that for the HCW, DI and CIM, the effect of covariates is taken into account through the outcomes of control units which the covariates affect.}. 
This can in turn lead to more precise estimates of the counterfactuals. 
Covariates that are potentially affected by the intervention should not be included when using any of the methods except the SCM, because the treatment-free values of the covariate are not observed in the post-intervention period.
The SCM can use these covariates in the pre-intervention period to estimate the weights.  
} 

\textcolor{black}{
There will be applications where more than one method is appropriate. 
This is to be expected considering their connections explained in Section \ref{sec:connections}. 
For example, the SCM, HCW, DI and CIM estimators are all well-suited when $T_1$ is large and there are few control units. 
In such cases, users might choose any of these methods. 
However, it is still worth applying the remaining methods in order to check that conflicting results are not obtained. 
Methods that perform poorly on diagnostic checks or are based on assumptions that seem unrealistic for the dataset of interest should not be considered. 
Even within the same method a sensitivity analysis is recommended. 
This can be carried out by implementing the method using different model specifications as explained in Section \ref{sec:model}. 
Ideally, results obtained from the different models should not conflict.  
For the SCM, HCW, DI and CIM, one can re-implement these methods excluding control units that received large coefficients (or weights for the SCM) in the first implementation, to provide reassurance that results are not driven by a single control unit.
}

\subsection{Connections with matching}
Our review does not cover matching methods even though some forms of matching are suitable for application in the setting that we are investigating. 
This is because we view the SCM as the best suited matching method in this setting: by using data on all control units it attempts to construct an exact match for the treated unit\footnote{HCW, DI and CIM also attempt to construct an exact match for the treated units. However, these methods may rely on extrapolation, which is not done in matching approaches.}.

\textcolor{black}{
However, matching can be used prior to applying the methods described in this paper, to restrict the pool of controls to those with similar characteristics to the treated units. 
This approach has been adopted for the DID \citep{ONeill2016}, LFM \citep{Gobillon2016} and CIM \citep{Schmitt2018}. 
For the DID method, \citet{Ryan2018} showed that matching can reduce biases that occur when the parallel trends assumption is violated. 
For a detailed overview of the matching literature in the context of causal inference with observational data, see \citet[Chapter 10]{Rosenbaum2002} or \citet{Stuart2010}. 
See \citet{Imai2018} for matching techniques for time-series data. 
}

\section{Proposals for future research}\label{sec:future}
There remain several open problems. 
Most existing methods do not fully account for autocorrelation in the outcome of the treated unit measured over time.  
In particular, the treatment effect estimates obtained by any of the methods except for the CIM are invariant to permutation of the time labels in the pre-intervention period. 
There may be potential gains in efficiency by extending these methods to account for structure over time.   

\textcolor{black}{
The SCM, HCW, DI and CIM assume a linear relationship between the outcome of the treated unit and the outcomes of control units but this is a strong assumption. 
\citet{Carvalho2018} account for non-linear relationships by regressing $y_{n_1+1,t}$ on transformations of the outcomes of control units but it is hard to choose which transformations to use. 
Therefore, it would be worth estimating the relationship between $y_{n_1+1,t}$ and the outcomes of the control units non-parametrically using, for example, machine learning techniques.
}

The methods we have described are designed to be applied to a single outcome. 
In the majority of applications there are several outcomes that may be affected by the intervention. 
For example, in the case study of Section \ref{sec:germany} we have considered per-capita GDP but there are alternative indexes, such as the unemployment rate, which we could instead be examined. 
Modelling of all outcomes jointly may provide a more precise estimate of the causal effect of intervention on any one of them. 
Although \citet{Robbins2017} provide an extension of the SCM method for multiple outcomes, the other methods could also benefit from being extended to handle multiple outcomes.

Another possible direction for future research is to develop models that take into account geographic location of units. 
In many applications, one might expect the outcomes on units with spatial proximity to be correlated. 
It would be useful to develop models that incorporate these correlations. 
\citet{Lopes2008} present a Bayesian LFM that models the correlation between the loadings of any two units as a function of the distance between these units. 
Their model could be used to estimate intervention effects with minor modifications.

We will investigate some of these problems in our future work.

%% file: appendix.tex
\appendix

\section{Unbiasedness and consistency}\label{sec:theory}

Unbiased or asymptotically unbiased estimates of $\tau_{it}$ can be obtained with all four methods described in this review. 
For each one of them, we now describe the sampling framework and the main assumptions for the unbiasedness to hold. 
For ease of exposition, we choose not to list some technical regularity conditions required for the results presented to hold; readers can refer to the original publications for these.

\textit{DID}. 
For the linear DID estimator, we make use of some well-known results for OLS regression, see e.g.\ \citet{Wool2013}. 
If the DID model of Equation \eqref{eq:didb} holds then $\hat\tau_{it}^\mathrm{DID}$ is unbiased, that is, 
\begin{equation}\nonumber
\mathbb{E}\left[\hat\tau_{it}^\mathrm{DID}\right]=\tau_{it},
\end{equation}
where the expectation is taken with respect to the conditional distribution of $\boldsymbol\varepsilon=\left(\varepsilon_{11},\dots,\varepsilon_{1T},\dots,\varepsilon_{n1},\dots,\varepsilon_{nT}\right)^\top$ given $\boldsymbol{S}_n$, where $\boldsymbol{S}_n=(\boldsymbol{x}_{11}^\top,\dots,\boldsymbol{x}_{1T}^\top,\kappa_1,\boldsymbol{d}_{1}^\top,\dots,\\ \boldsymbol{x}_{n1}^\top,\dots,\boldsymbol{x}_{nT}^\top,\kappa_n,\boldsymbol{d}_{n}^\top)^\top$. 
That is, $\boldsymbol{S}_n$ is common to the repeated samples but the errors $\boldsymbol{\varepsilon}$ differ in each repeated sample.

\textit{LFM}. 
\citet{Xu2017} study the properties of $\hat\tau^{\mathrm{XU}}_{it}$. 
If the LFM of Equation \eqref{eq:factor} holds then under some regularity conditions (which include weak serial correlation of the error terms within each unit) $\hat\tau^{\mathrm{XU}}_{it}$ is asymptotically unbiased, that is
\begin{equation}
\nonumber \mathbb{E}\left[\hat\tau^{\mathrm{XU}}_{it}\right]\rightarrow \tau_{it} 
\end{equation}
as $n_1\rightarrow\infty$ and $T_1\rightarrow\infty$, where the expectation is taken with respect to the conditional distribution of $\boldsymbol{\varepsilon}$ given $\boldsymbol{S}_n$, where $\boldsymbol{S}_n=(\boldsymbol{x}_{11}^\top,\dots,\boldsymbol{x}_{1T}^\top,\boldsymbol\lambda_1^\top,\boldsymbol{d}_{1}^\top,\dots,\boldsymbol{x}_{n1}^\top,\\\dots,\boldsymbol{x}_{nT}^\top,\boldsymbol\lambda_n^\top,\boldsymbol{d}_{n}^\top,\boldsymbol{f}_1^\top,\dots,\boldsymbol{f}_T^\top)^\top$. 
Intuitively, we require that both $n_1$ and $T_1$ are large in order to accurately estimate factors at each post-intervention time point and loadings for the treated units, respectively, which we need in order to predict the counterfactual outcomes.


\textit{Synthetic control-type approaches}. 
These methods do not assume a generative model, but rather exploit linear relationships between the data on the treated and control units in order to construct a counterfactual. 
Such relationships may arise from various data-generating mechanisms. 
Hence, their unbiasedness properties can be studied under any of these. 

Assume that the LFM
 \begin{eqnarray}\label{eq:sclfm}\nonumber
 \quad y_{it}&=&y_{it}^{(0)}+\tau_{it} d_{it} \\ 
  y_{it}^{(0)}&=&\mu_t + \boldsymbol{x}^\top_{i}\boldsymbol\theta_t + \boldsymbol\lambda_i^\top\boldsymbol{f}_t  +\varepsilon_{it},
 \end{eqnarray}
holds, the error terms $\varepsilon_{it}$ have zero mean given $\boldsymbol{S}_n$ and $\varepsilon_{it}\independent\varepsilon_{js}$ given $\boldsymbol{S}_n$, ($i\neq j$ and $t\neq s$), where $\mu_t$ are time fixed-effects, $\boldsymbol{x}_i=(x_{i1},\ldots,x_{iK})^\top$ are time-invariant covariates and $\boldsymbol{S}_n=(\boldsymbol{x}_{1}^\top,\boldsymbol\lambda_1^\top,\boldsymbol{d}_{1}^\top,\dots,\boldsymbol{x}_{n}^\top,\boldsymbol\lambda_{n}^\top,\boldsymbol{d}_{n}^\top,\boldsymbol{f}_1^\top,\dots,\boldsymbol{f}_T^\top,\mu_1,\ldots,\\\mu_t)^\top$. 
\citet{Abadie2010} show that if there exist $\varpi_1,\dots,\varpi_{n_1}$ such that
\begin{eqnarray}\label{eq:scvarpi}\nonumber
	\boldsymbol{\lambda}_{n_1+1}& = & \sum_{i=1}^{n_1}{\varpi_i\boldsymbol{\lambda}_{i}}
	\\ 
	\boldsymbol{x}_{n_1+1}& = & \sum_{i=1}^{n_1}{\varpi_i\boldsymbol{x}_{i}},
\end{eqnarray}
then under some regularity conditions $\hat\tau_{n_1+1,t}^\mathrm{SC}$ is asymptotically unbiased, \textit{i.e.}
\begin{equation}\nonumber 
\mathbb{E}\left[\hat\tau_{n_1+1,t}^{\mathrm{SC}}\right]\rightarrow \tau_{n_1+1,t}
\end{equation}
as $T_1\rightarrow \infty$, where the expectation is taken with respect to the conditional distribution of $\boldsymbol\varepsilon$ given $\boldsymbol{S}_n$. 
The conditions \eqref{eq:scvarpi} imply that both observed ($\boldsymbol{x}_{n_1+1}$) and unobserved ($\boldsymbol{\lambda}_{n_1+1}$) characteristics of the treated unit lie in the convex hull of the characteristics of control units, thus allowing interpolation.
When this is not true (\textit{i.e.}\ when such $\varpi_1,\dots,\varpi_{n_1}$ do not exist, thus forcing extrapolation to be used), the SCM estimator will be generally biased \citep{Gobillon2016,Ferman2016a}. 

Assume the following variant of the LFM: 
\begin{eqnarray}\label{eq:hcw}
  y_{it}^{(0)}&=&\kappa_i+\boldsymbol\lambda_i^\top\boldsymbol{f}_t +\varepsilon_{it},
\end{eqnarray}
where $\kappa_i$ are unit fixed effects and $\varepsilon_{it}$ are zero-mean, homoscedastic error terms which are independent of $\boldsymbol{f}_s$ for all $t,s$ and independent of $d_{js}$ for all $i\neq j$. 
\citet{Hsiao2012} prove that if there exist $\gamma_1,\dots,\gamma_{n_1}$ such that
\begin{equation}\label{eq:hcw1}
\lambda_{n_1+1,j} = \sum_{i=1}^{n_1}\gamma_i\lambda_{ij}
\end{equation}
is true for every $j=1,\dots,J$ (along with some technical conditions), then $\hat{\tau}_{n_1+1,t}^\mathrm{HCW}$ is unbiased, \textit{i.e.}
\begin{equation}
\nonumber \mathbb{E}\left[\hat{\tau}_{n_1+1,t}^\mathrm{HCW}\right] = \tau_{n_1+1,t},
\end{equation}
where the expectation is taken with respect to the conditional distribution of  $\boldsymbol\varepsilon$ given $\boldsymbol{S}_n$, where $\boldsymbol{S}_n=(\kappa_1,\boldsymbol{\lambda}_1^\top,\boldsymbol{d}_1^\top,\dots,\kappa_{n},\boldsymbol{\lambda}_{n}^\top,\boldsymbol{d}_{n}^\top,\boldsymbol{f}_1^\top,\dots,\boldsymbol{f}_T^\top)^\top$.

\textit{CIM}. 
If the CIM of Equations \eqref{eq:bstseg} is the true data-generating model and the prior on the vector of model parameters $\boldsymbol\vartheta=(\beta_1,\ldots,\beta_{n_1},\sigma_\varepsilon^2,\sigma_\eta^2,\sigma_\zeta^2)^\top$ assigns non-zero probability to its true value, then the posterior distribution of $\boldsymbol{\vartheta}$ will converge to a point mass on its true value as $T_1\rightarrow\infty$. 
Consequently, the posterior mean of $y^{(0)}_{n_1+1,t}$ will converge to its true value, and so $\hat{\tau}_{n_1+1,t}^\mathrm{CIM}$ is an asymptotically unbiased estimate of of $\tau_{n_1+1,t}$ (as $T_1\rightarrow\infty$).

The above results concern (asymptotic) unbiasedeness. 
Consistent estimation of $\tau_{it}$ is not feasible, unless it is assumed that $\tau_{it} = \tau_i$ or $\tau_{it} = \tau_t$, \textit{i.e.} that the unit-specific treatment effects are the same at all post-intervention times or (when $n>n_1+1$) are the same at each time for all treated units. 
This is because, regardless of how many units and timepoints there are, $y_{it}^{(1)}$ is only measured once for each $i>n_1$ and $t>T_1$. 
It is not uncommon to assume that $\tau_{it}=\tau$ (\textit{e.g.} \citet{Angrist2009,Gobillon2016}). When this is done, existing results for the linear DID model (\textit{e.g}.\ \citet{Wool2013}) and the LFM method of \citet{Xu2017} (\textit{e.g}.\ \citet{Bai2009}) imply consistency of the estimator of $\tau$ when either of those methods are used. 
These results, though, require some additional technical assumptions to hold. 
Alternatively, a looser structure could be imposed on $\tau_{it}$. For example, HCW assume that $\tau_{n_1+1,T_1+1}, \ldots, \tau_{n_1+1,T},$ is an auto-regressive moving-average process. 
This enables the mean of this process to be consistently estimated.

\section{Real data supplementary analysis}\label{sec:suppl} 
\begin{figure}[htp]
        \centering
        \includegraphics[scale=0.4]{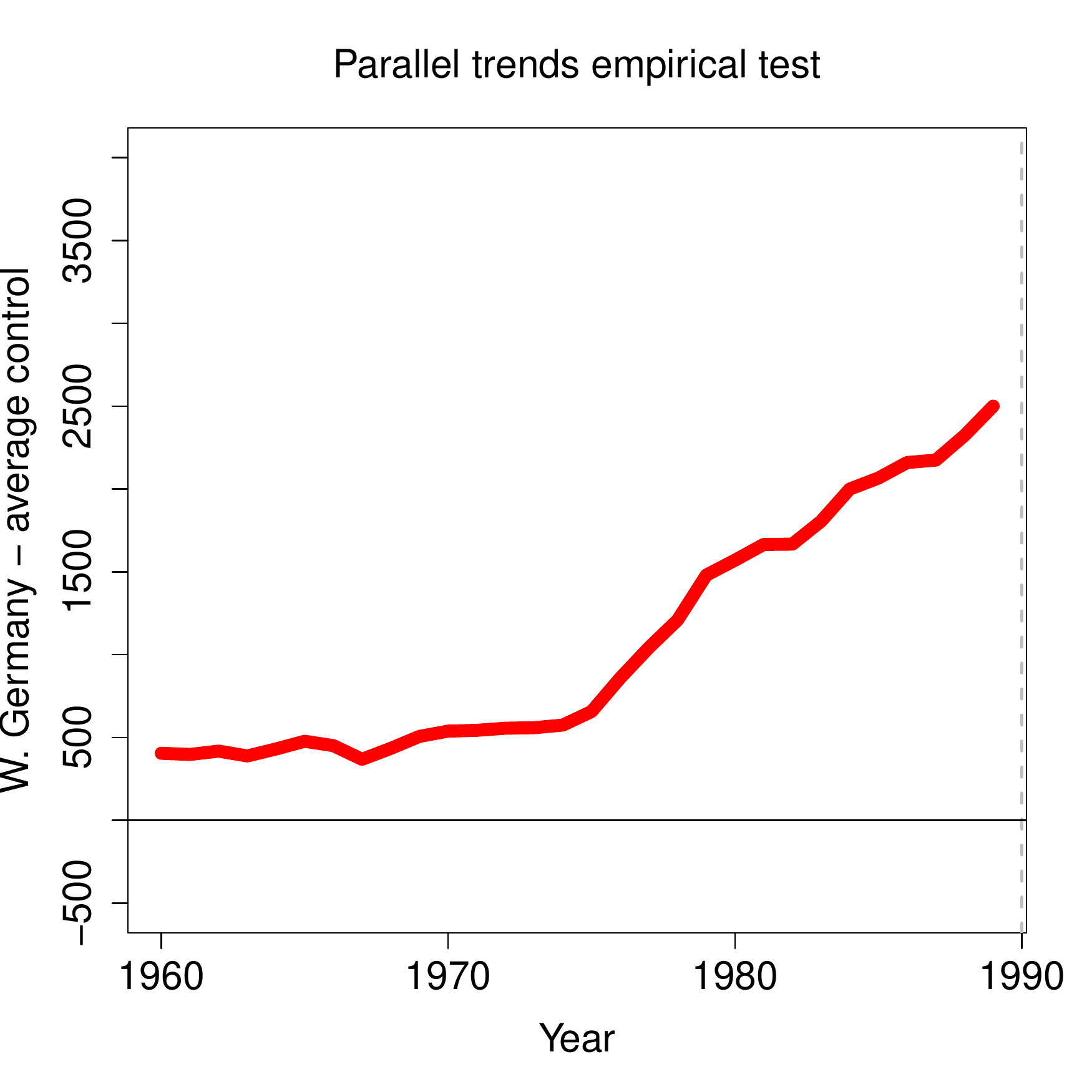}
        \vspace{-0.15in}
      \caption{Difference over time between West Germany's per capita GDP and the average of control countries. Rather than being constant, the difference increases over time thus suggesting that the DID parallel trends assumption might not be plausible.}      
      \label{fig:didpar}
 \end{figure}

In this section, we provide supplementary material for the data analysis of Section \ref{sec:real1}. 
Figure \ref{fig:didpar} shows the parallel trends diagnostic check described in Section \ref{sec:blacks}.
The estimated factor loadings obtained by applying the LFM method of \citet{Xu2017} are shown in Table \ref{tab:loadings}.
Finally, the $r$ statistics obtained by applying the empirical test of \citet{Abadie2015} with the SCM and HCW methods are shown in Table \ref{tab:rstat}.  

 \begin{table}[ht]
 \centering
\begin{tabular}{rrrrrrrrrrr}
  & \multicolumn{10}{c}{\textbf{Factor}}\\
\textbf{Country} & 1 & 2 & 3 & 4 & 5 & 6 & 7 & 8 & 9 & 10 \\ 
  \hline\hline
West Germany & -0.75 & -0.16 & 0.29 & 0.26 & -0.15 & -0.12 & -0.80 & 0.35 & -0.42 & -1.42 \\ 
  Australia & -0.14 & 0.47 & 0.91 & -0.55 & 1.71 & -0.62 & -0.34 & -0.64 & -0.25 & 0.93 \\ 
  Austria & -0.50 & -0.52 & -0.17 & 0.18 & -0.69 & -1.06 & 0.26 & 0.45 & -2.28 & 0.21 \\ 
  Belgium & -0.22 & -0.46 & 0.10 & 1.02 & -0.44 & -0.32 & 0.98 & 1.61 & -0.43 & -0.20 \\ 
  Denmark & -0.34 & 0.24 & 0.21 & -1.34 & 0.09 & -1.01 & -0.93 & -0.25 & 1.24 & 0.02 \\ 
  France & -0.14 & -0.33 & 0.05 & 0.11 & -0.45 & -0.18 & -0.95 & 1.12 & -0.19 & -0.92 \\ 
  Greece & 1.95 & -0.02 & 1.52 & 1.10 & -0.55 & -0.88 & 0.17 & -1.99 & -0.01 & -1.54 \\ 
  Italy & -0.03 & -0.72 & -0.62 & -0.65 & -0.39 & -0.26 & -0.49 & 0.48 & -0.85 & -1.14 \\ 
  Japan & -0.33 & -1.18 & -2.04 & -0.30 & -0.07 & -0.53 & 2.30 & -1.28 & 1.15 & 0.12 \\ 
  Netherlands & -0.45 & 0.31 & 0.02 & 1.30 & 0.07 & -1.68 & -0.83 & 1.08 & 1.99 & 0.74 \\ 
  New Zealand & 1.32 & -0.12 & 1.40 & -2.35 & -0.26 & 0.35 & 1.24 & 1.10 & 0.14 & 0.96 \\ 
  Norway & -1.60 & 2.55 & 0.29 & -0.15 & -2.04 & 0.83 & 0.53 & -0.70 & 0.08 & -0.05 \\ 
  Portugal & 1.73 & 0.49 & -2.18 & -0.58 & -0.53 & 0.92 & -1.74 & -0.17 & 0.02 & 0.14 \\ 
  Spain & 0.98 & 0.85 & -0.44 & 1.57 & 0.74 & 0.54 & 0.37 & -0.16 & -0.93 & 2.16 \\ 
  Switzerland & -0.71 & -2.19 & 1.10 & 0.70 & -0.71 & 2.23 & -0.77 & -0.49 & 0.77 & 0.77 \\ 
  UK & 0.10 & 0.89 & -0.10 & 0.49 & 1.92 & 1.58 & 0.88 & 1.04 & 0.61 & -1.92 \\ 
  USA & -1.63 & -0.24 & -0.05 & -0.55 & 1.58 & 0.07 & -0.68 & -1.20 & -1.05 & -0.28 \\ 
   \hline\hline
\end{tabular}
 \caption{Factor loadings for the 17 countries, as obtained by fitting the LFM of \citet{Xu2017} to the West German reunification data.}
 \label{tab:loadings}
 \end{table}

  \begin{table}[ht]
  \centering
  \begin{tabular}{lrr}
     & \multicolumn{2}{c}{\textbf{Method}} \\
  \textbf{Country} & \textbf{SCM} & \textbf{HCW}\\ 
    \hline \hline
 West Germany & 30.72 & 71.53 \\ 
 Australia & 6.86 & 16.83 \\ 
    Austria & 4.24 & 45.04 \\ 
    Belgium & 4.52 & 16.90 \\ 
    Denmark & 6.25 & 21.35 \\ 
    France & 8.00 & 52.07 \\ 
    Greece & 7.83 & 18.72 \\ 
    Italy & 20.48 & 46.72 \\ 
    Japan & 4.87 & 29.73 \\ 
    Netherlands & 20.44 & 36.83 \\ 
    New Zealand & 5.16 & 16.99 \\ 
    Norway & 13.78 & 77.36 \\ 
    Portugal & 0.70 & 57.69 \\ 
    Spain & 7.53 & 14.32 \\ 
    Switzerland & 2.36 & 29.75 \\ 
    UK & 6.94 & 30.27 \\ 
    USA & 5.97 & 42.55 \\ 
     \hline\hline
  \end{tabular}
  \caption{$r$ statistics obtained by applying the empirical test of \citet{Abadie2015} to the German reunification data, for both the SCM and HCW methods.}
  \label{tab:rstat}
  \end{table}   

